\documentclass[twocolumn,showpacs,preprintnumbers,amsmath,amssymb,superscriptaddress,groupedaddress]{revtex4-1}

\usepackage{graphics}
\usepackage{epsfig}

\begin{document}

\title{Reconfigurable nonreciprocity with nonlinear Fano diode}

\author{Yi Xu$^{1}$ and Andrey E. Miroshnichenko$^{2}$}
\email{andrey.miroshnichenko@anu.edu.au}
\affiliation{$^{1}$School of Physics and Electronic Engineering, Guangzhou University, Guangzhou 510006, P.R. China}
\affiliation{$^{2}$Nonlinear Physics Centre, Research School of Science and Engineering, Australian National University, Canberra ACT 0200, Australia}

\begin{abstract}
We propose a dynamically tunable nonreciprocal response for wave propagations by employing nonlinear Fano resonances. We demonstrate that transmission contrast of waves propagation in opposite directions can be controlled by excitation signal. In particular, the unidirectional transmission can be flipped at different times of a pulse, resembling a diode operation with {\em dynamical reconfigurable nonreciprocity}. The key mechanism is the interaction between the linear and nonlinear Fano resonances that allows for the tunable unidirectional wave propagation and ultrahigh transmission contrast ratio. We further present a realistic photonic example which demonstrates the properties of nonreciprocity can be dynamically manipulated using a pump pulse, based on the general theoretical model. 
\end{abstract}
\maketitle

\section{Introduction}
Electronic diodes are one of the key elements in modern electronic devices, and even play an important role in our daily lives. The most common function of a diode is to allow an electric current to pass in one direction, while blocking current in the opposite direction, thus enabling the unidirectional current flux and the rectification of an electrical signal. To date, many contributions have been made for the rectification of different types of energy flows. For example, thermal diode for thermal flow [See Ref.~\cite{thermal} and references therein], acoustic diode for sonic wave~\cite{acoustic,asym} and electromagnetic diode for electromagnetic wave~\cite{optics1,sfm,tunable_nature,left_hand,lin,e_diode,on-chip,hu,diode_passive} have been demonstrated both theoretically and experimentally. As a particular example, violation of the Lorentz reciprocity in optical system prevents the light from retracing its directional transmission and, thus, facilitates certain potential applications, such as optical diodes and isolators~\cite{comment_fan}. There are several mechanisms which allow to break the reciprocity, including magneto-optical effect ~\cite{time,DJ,on-chip,davoyan}, nonlinear material~\cite{optics1,sfm,left_hand,lin,e_diode,hu,diode_passive} and structure with time-dependent refractive index ~\cite{fan_theory,fan_experiment}. High-speed and on-chip optical circuit would require all-optical modulation, in which using nonlinear optical effect of the system could be one of the solutions.~\cite{diode_passive}. 

\begin{figure}[htb]
\includegraphics[width=\columnwidth]{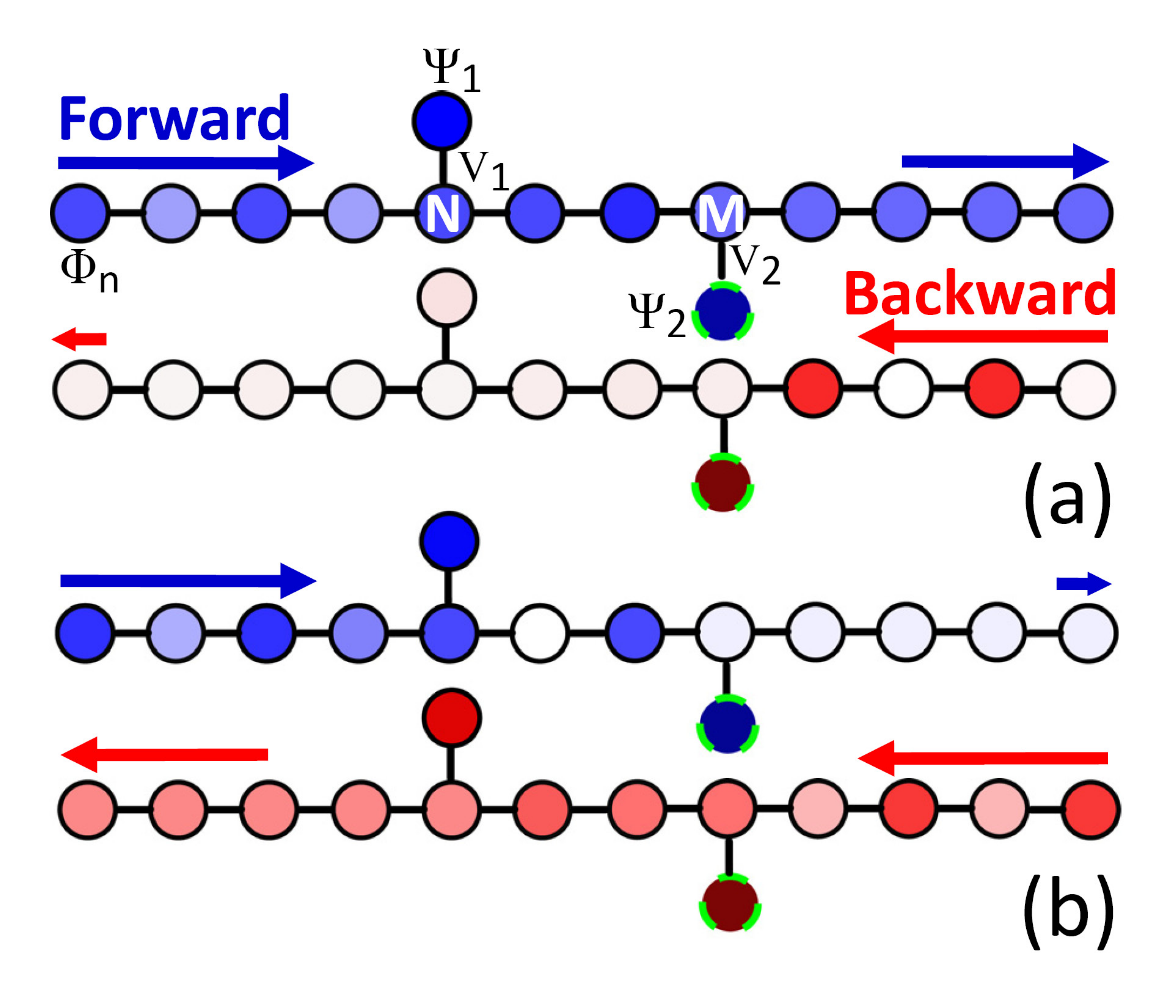}
\caption{\label{fig:fig1} 
(Color online) Schematic view of the reversible nonreciprocal response based on the nonlinear Fano resonance. The arrows indicate the direction of the forward and backward transmission. The same system can support waves propagation predominantly either in one (a) or opposite  (b) direction for different excitation signals. The color scale stands for the intensity of the field in each cite which are two particular solutions of Eq. 1. Nonlinear Fano defect is shown by a dash ring while others are linear.
}
\end{figure}

Reconfigurable light-driven isolator attracts much attention due to its capacity of switching on/off the interband transition, which manifests itself as the transition from nonreciprocal to reciprocal response, by using a control light beam~\cite{Russell}. Recently, an interesting phenomenon named {\em reversible optical nonreciprocity} was proposed where a nonreciprocal response can be flipped from transmitting a signal predominantly in one direction to opposite one~\cite{aem_non}. However, the contrast of such system is low due to the limited scattering channel. And the possibility of dynamical manipulation has not been addressed yet. By offering such possibilities for dynamically tuning of the nonreciprocity, one can direct the energy flow in real-time and thus realize more advanced control of the wave propagating systems. For example, diodes with tunable forward and backward transmission rates were proven to be significant in manipulating the properties of the wave rectification~\cite{tunable_nature}. Furthermore, if the properties of nonreciprocity can be controlled instantaneously, a lot of basic wave propagation devices, such as switch, router and rectifier etc., could be realized by employing this effect. Therefore, realizing the reconfigurable function of nonreciprocity is very crucial for precise wave manipulation, as the demonstration of reconfigurable photonic crystal (PhC) circuits~\cite{Ben}. 
\begin{figure}[htb]
\includegraphics[width=\columnwidth]{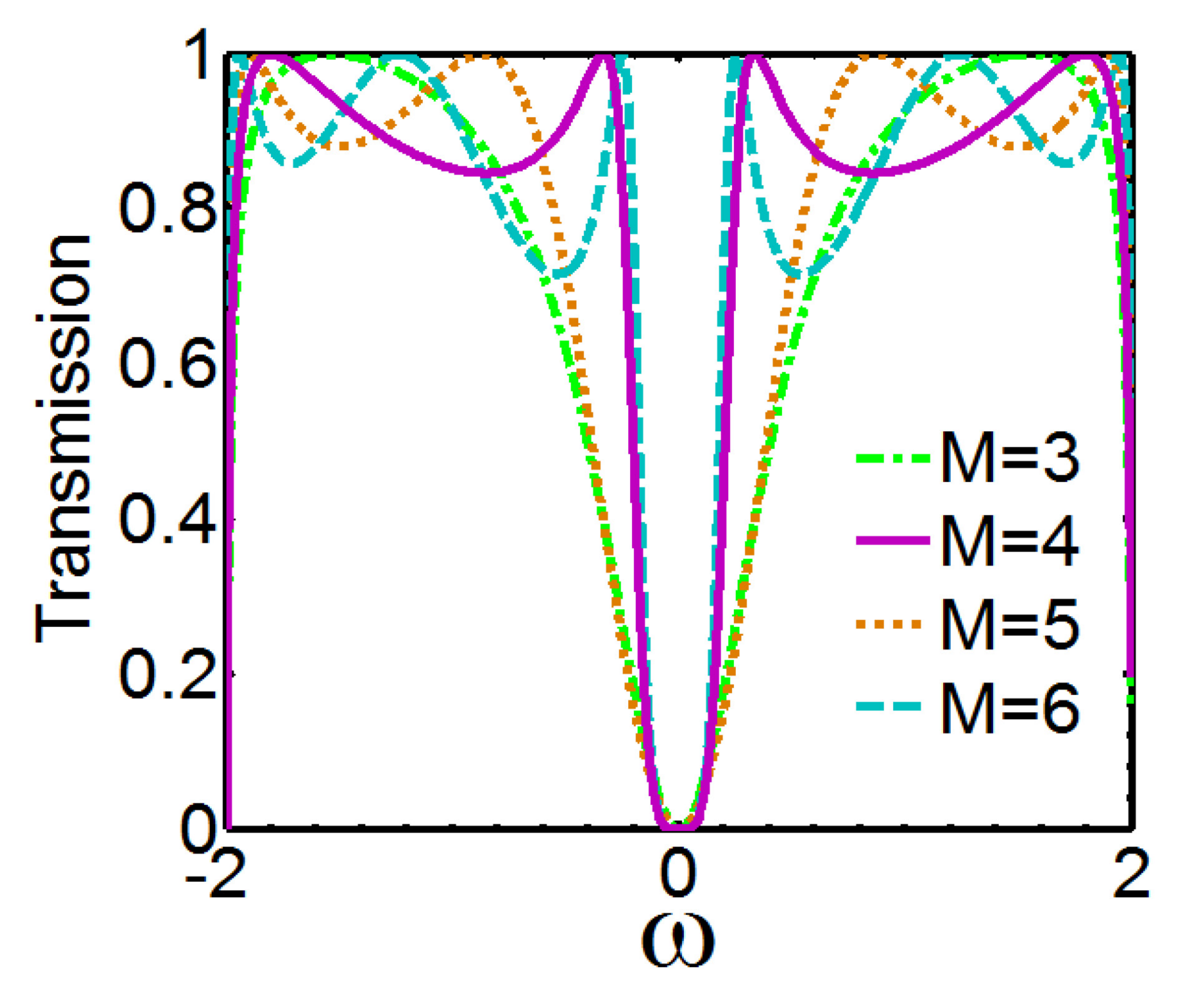}
\caption{\label{fig:fig2} 
(Color online) Linear transmission of the system presented in Fig.~\ref{fig:fig1}. Without loss of general, we put here $ N=1 $, $ E_{d1,d2}=0 $ and $ V_{1,2}=0.6 $.
}
\end{figure}

One of the best ways for achieving the high transmission contrast can rely on the Fano resonances, where the transmission can completely vanish due to destructive interference~\cite{fan,aem_rev,B_review,LPR}. Thus, diodes based on the Fano resonances may benefit from their typical resonant reflection and thus ultrahigh-contrast-ratio~\cite{ultra}. It was shown that coupling two nonlinear Fano defects can induce the symmetry breaking of the system~\cite{maes}. Such system reveals rich nonlinear dynamics and thus facilitates one's manipulation of the optical bistable scattering~\cite{bound1,bound2}. In this paper, we suggest that an even simpler system, consisting of a waveguide with two side-coupled Fano defects, one is with cubic nonlinearity while the other one is linear, exhibits quite interest physical property, namely, {\em dynamical reconfigurable nonreciprocity} (DRN). We demonstrate that the interaction between a linear and a nonlinear Fano resonances has a pronounced DRN. It is shown that the unidirectional transmission of nonreciprocity can be further manipulated not only by choosing the operating frequency and geometry parameters, but also dynamically with the input signal. 

\section{Model}
The transmission properties of proposed nonreciprocal system shown in Fig.~\ref{fig:fig1} can be studied by using the modified Fano-Anderson model~\cite{aem1}. It allows us to derive analytical solutions of the nonlinear transmission at reversal incident directions. The equations describing the nonlinear dynamics of the scattering are:
\begin{eqnarray}
\label{eq:eq1}
&i\dot{\phi}_n&=C(\phi_{n-1}+\phi_{n+1})+\delta_{n,N} V_{1}\psi_{1}  +\delta_{n,M} V_{2}\psi_{2}  \nonumber\\
&i\dot{\psi}_{1}&=V_{1}\phi_{N}+E_{d1}\psi_{1}   \nonumber\\
&i\dot{\psi}_{2}&=V_{2}\phi_{M}+E_{d2}\psi_{2}+\lambda\vert\psi_{2}\vert^{2}  \nonumber\\
\end{eqnarray}
where the overdot stands for the derivative in time, $ \phi_{n} $ and $ \psi_{1,2} $ represent the complex fields of the sites in the chain and side coupled defects, respectively. $ E_{d1,d2} $ is the defect energy, $ \lambda $ is the nonlinearity parameter, $ C $ is the nearest-neighbour coupling constants, $ \delta_{nm} $ is the Kronecker delta symbol and $ V_{1,2} $ is the side-coupled strength between defects and the chain. The transmission matrix for the system, where $ \varepsilon_{1,2}=V^{2}_{1,2}/(\omega-E_{d1,d2}) $, $ \omega=2C\cos q $ and $ C_{q}=2C\sin q $, is shown in Eq.~\ref{eq:eq2} with $ \phi_{n}=A_{n}e^{-i\omega t} $ and $ \psi_{1,2}=B_{1,2}e^{-i\omega t} $, $ A_{n} $ and $ B_{1,2} $ are complex numbers~\cite{aem1}. 
 
\begin{eqnarray}
\label{eq:eq2}
K=\dfrac{-1}{C_{q}^{2}}\left[ 
  \begin{array}{ccc}
  \varepsilon_{1}+iC_{q}&\varepsilon_{1}e^{-i2Nq}\\
  -\varepsilon_{1}e^{i2Nq}&-\varepsilon_{1}+iC_{q}\\
  \end{array}
\right] 
\left[ 
  \begin{array}{ccc}
  \varepsilon_{2}+iC_{q}&\varepsilon_{2}e^{-i2Mq}\\
  -\varepsilon_{2}e^{i2Mq}&-\varepsilon_{2}+iC_{q}\\
  \end{array}
\right] \nonumber\\
\end{eqnarray}
Linear transmission of the system can be obtained by $ T=\vert 1/K(2,2)\vert^{2} $. Four representative examples are plotted in Fig.~\ref{fig:fig2}. As can be seen from these results, when $ M-N=2n+1 $ and $ n=1,2,3\cdots $ the transmission forms a nearly flat bottom stop band with sharp edge near the Fano resonances of the side-coupled defects. 

\begin{figure}[htb]
\includegraphics[width=\columnwidth]{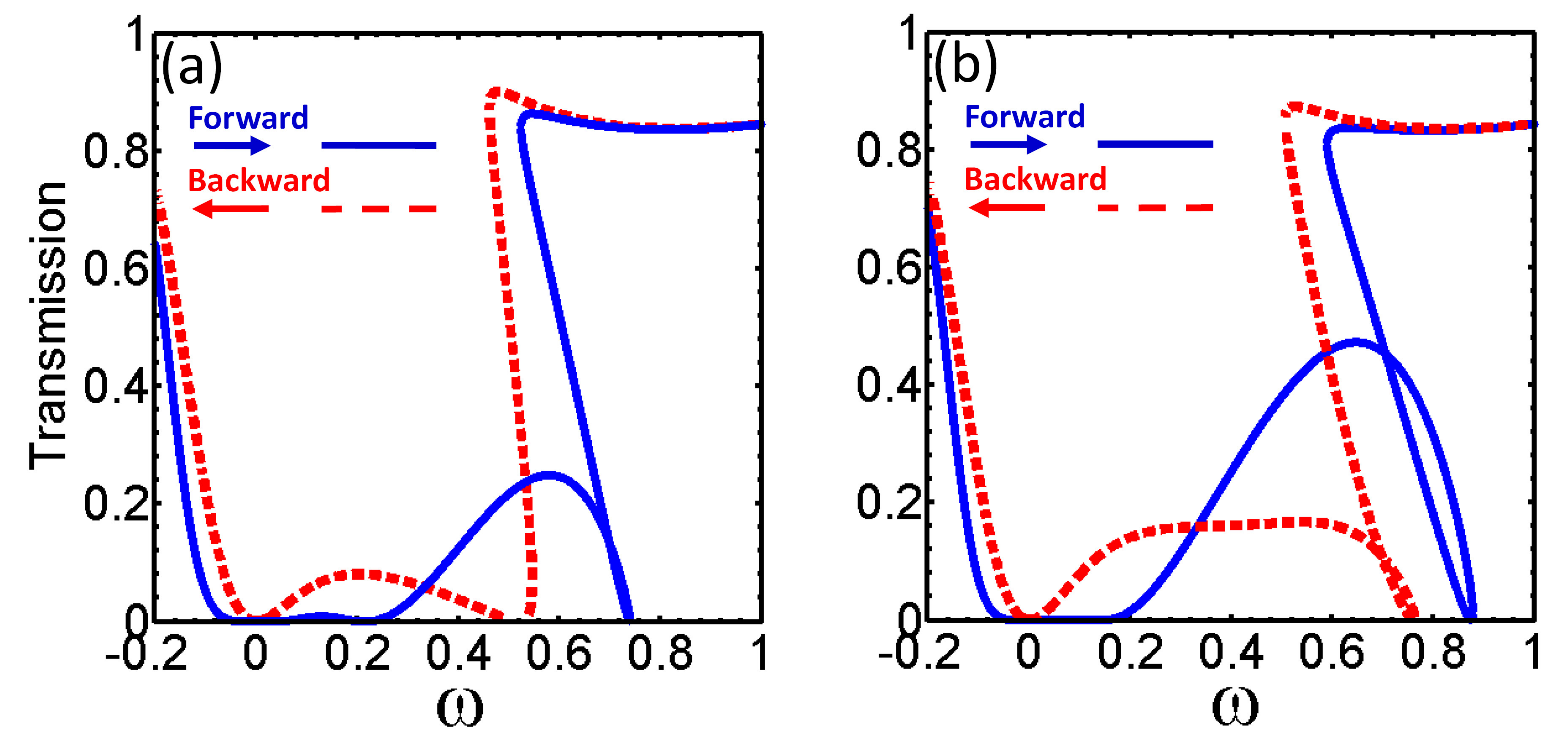}
\caption{\label{fig:fig3} 
(Color online) (a) Nonlinear transmission of opposite incident directions. Here $ N=1 $, $ M=4 $, $ E_{d1,d2}=0 $, $ V_{1,2}=0.6 $, $ \lambda=1 $ and incident power $ \vert I\vert^{2}=0.05 $. (b) The same plot as (a) except for $ \vert I\vert^{2}=0.08 $. Blue solid/dashed red color indicates the forward/backward incident direction throughout this paper, as is indicated in Fig. 1 and Fig. 3.
}
\end{figure}

When we add a cubic nonlinearity to one of the Fano defects, above certain threshold the reciprocity can be broken. Moreover, the indirect interaction between the linear and nonlinear Fano defects via the chain suggests the possibility of the {\em reversible nonreciprocal effect}~\cite{aem_non}. In particular, incident wave propagating in opposite directions would excite the nonlinear Fano defect with different rates, as is shown in Fig.~\ref{fig:fig1}. Because the nonlinear Fano resonance is the input power dependent~\cite{aem1}, different power injection would lead to distinct shift of the nonlinear Fano resonance, which, in turn, would interact differently with the linear Fano resonance. Furthermore, the bistability of the nonlinear Fano resonance offers another opportunity to control the interaction by dynamically choosing different branches in the hysteresis loop of the bistable state~\cite{aem1}. Such tunable interaction gives rise the possibility for dynamically manipulating the physical properties of wave scattering. The nonlinear transmission of the system can be written as follow

\begin{figure*}[!t]
\includegraphics[width=2\columnwidth]{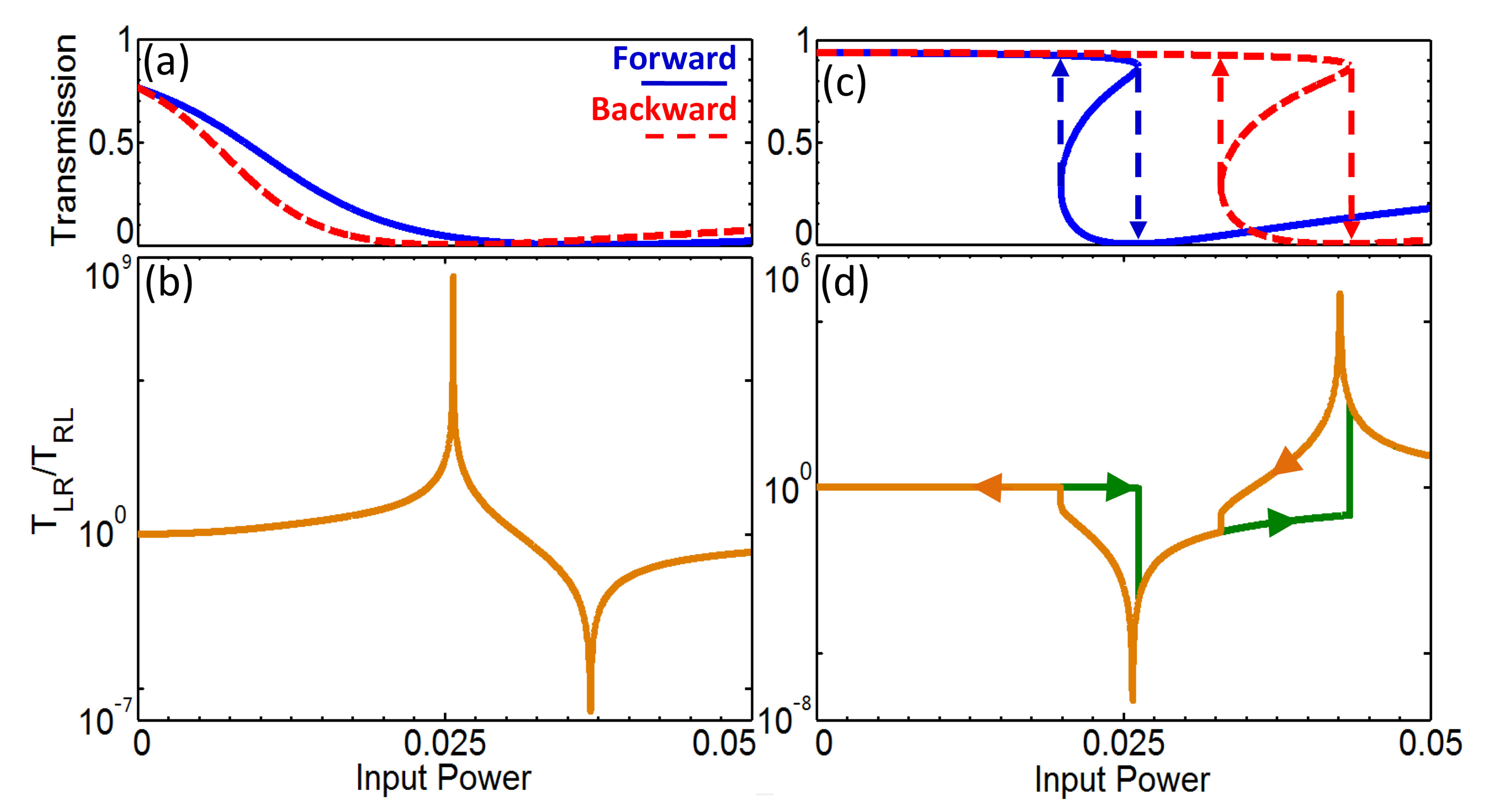}
\caption{\label{fig:fig4} 
(Color online) (a) The dependence of nonlinear transmission on the input power of the system for opposite incident directions (frequency $ \omega=0.28 $). Here $ N=1 $, $ M=4 $, $ E_{d1,d2}=0 $, $ \lambda=1 $ and $ V_{1,2}=0.6 $. Solid lines represent the forward excitation while dashed lines indicate the backward one. (b) The corresponding $ T_{LR}/T_{RL} $ contrast-ratio versus input power in (a). (c) and (d) The same plots as (a) and (b) except for $ \omega=0.45 $. (a)-(d) share similar $ x $ label. The arrows indicate the transition directions between two stable states.
}
\end{figure*}
\begin{widetext}
\begin{eqnarray}
\label{eq:eq3}
T_{LR/RL}=\dfrac{\alpha_{q1}^{2}x^{2}}{2(1-\cos 2ql)(1-\alpha_{q1}x)+\alpha_{q1}^{2}x^{2}-2\sin(2ql)(\alpha_{q1}+x)+(\alpha_{q1}+x)^{2}},\nonumber\\
\end{eqnarray} 
\end{widetext}
where $ \alpha_{q1}=C_{q}(\omega-E_{d1})/V_{1}^{2} $, $l=M-N$ and $ x=-\cot \theta $, $ \theta $ is the scattering phase of the nonlinear Fano defect. While $ x $ is the real solutions of the following cubic equation for the elastic scattering problem~\cite{QM}:
\begin{eqnarray}
\label{eq:eq4}
ax^{3}+bx^{2}+cx+d_{LR/RL}=0
\end{eqnarray}
with $ a=\alpha_{q1}^{2}+1$, $ b=(2\cos 2ql+1)\alpha_{q1}-2\sin 2ql-\alpha_{q2}(\alpha_{q1}^{2}+1) $, $ c=2-2\cos 2ql-2\alpha_{q1}\sin 2ql+\alpha_{q1}^{2}-2\alpha_{q2}(\alpha_{q1}\cos 2ql-\sin 2ql)$, $ d_{LR}=-\alpha_{q2}\gamma+\lambda C_{q}^{3}\alpha_{q1}^{2}I^{2}/V_{2}^{4} $ and $ d_{RL}=-\alpha_{q2}\gamma+\lambda C_{q}^{3}\alpha_{q1}^{2}I^{2}\eta/V_{2}^{4} $, $ \gamma=2-2\cos 2ql-2\alpha_{q1}\sin 2ql+\alpha_{q1}^{2} $ and $ \eta=\vert e^{3iq}-\omega V_{1}^{2}+V_{1}^{2}/\omega\vert^{2} $  for the case $ l=3 $ and $ C=1 $. Here LR means the forward transmission (blue color) and RL denotes the backward case (red color) as are indicated in Fig. 1 (a).

As can be seen from the Eq.~\ref{eq:eq4}, the asymmetry scattering of the system arises when $ d_{LR}\neq d_{RL}$ and thus different roots of the cubic equation. Intuitively, such asymmetry ($ \eta\neq 1 $) can be fulfilled by choosing suitable frequency $ \omega $ and coupling strength $ V_{1} $. Fig.~\ref{fig:fig2} (a) and (b) are two examples with different input powers $ \vert I\vert^{2}=0.05 $ and $ \vert I\vert^{2}=0.08 $, respectively. Other parameters can be found in the caption. It can be seen from these figures that the system exhibits asymmetric response for opposite incident directions. Furthermore, given certain $ \omega $ and $ V_{1} $ preserving $ \eta\neq 1 $, the input power $I^2 $ dependence of both $ d_{LR}$ and $ d_{RL}$ offers us another degree of freedom to realize advanced manipulation of the nonreciprocity. When $ d_{LR}\neq d_{RL}$, the nonlinear Fano resonances of the system occur at different input power for reversal incident directions. Therefore, ultrahigh-contrast-ratio Fano diodes with reversible nonreciprocity are presented at these two corresponding transmission dips, as shown in Fig.~\ref{fig:fig4} (a)-(d). Fig.~\ref{fig:fig4} (a) presents the dependence of the forward and backward transmissions on the input power when $ \omega=0.28 $. There are distinct transmissions for reversal incident waves. Fig.~\ref{fig:fig4} (b) shows the corresponding $ T_{LR}/T_{RL} $ in Fig.~\ref{fig:fig4} (a). These results show that we can flip the unidirectional transmission by simply controlling the input power. Fig.~\ref{fig:fig4} (c) and (d) show the case when we use suitable frequency detuning $ \omega=0.45 $, i.e. different $ \eta $, to trigger bistability. Changing the input power is corresponding to tune the interaction between the linear and nonlinear Fano resonance into distinct bistable state and thus the rectification capacity of the system. It is demonstrated that the input power dependence of $ T_{LR}/T_{RL} $ can be engineered from tunable single value to tunable bistability by choosing different $ \eta $, which would facilitate the advanced control of the rectification, as are shown in Fig.~\ref{fig:fig4} (b) and (d).

\begin{figure}[htb]
\includegraphics[width=\columnwidth]{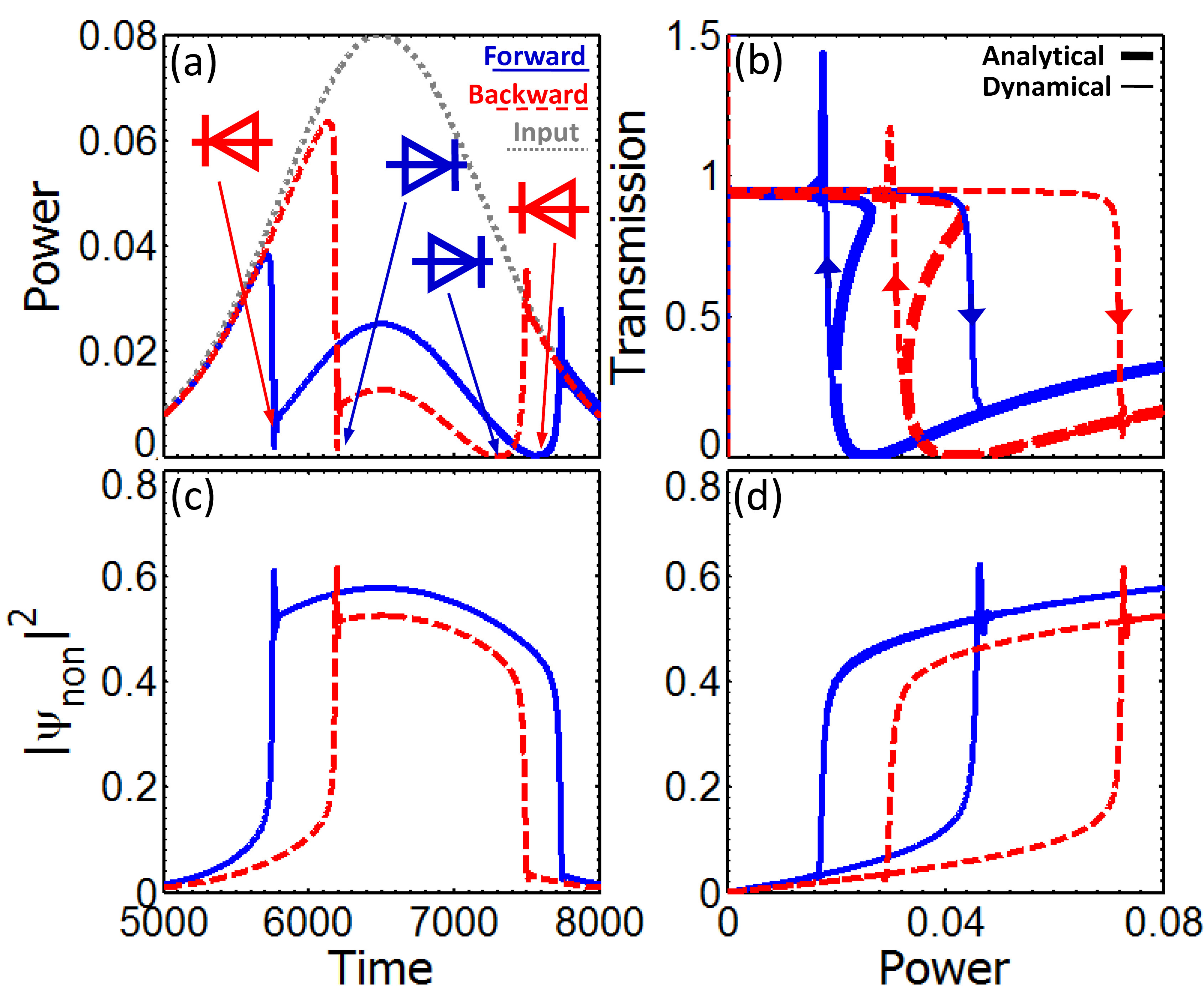}
\caption{\label{fig:fig5} 
(Color online) (a)Dynamics of the system at reversal incident direction with a Gaussian pulse with $ I=I_{0}\exp(-(t-t_{0})^{2}/W^{2})\sin(\omega t) $, where $ W=1400 $ and $ \omega=0.14 $. The geometry parameters are similar with the one in Fig.~\ref{fig:fig4} (c). The forward directions of the diode at specified pulse times are indicated by the insets. (b)Transmissions (thin lines) derived from (a) and the corresponding analytical results (thick lines). (c)Time evolutions of the nonlinear cavity excitations. (d)The effective pumping rates of the cavity. Solid lines stand for the forward excitation while dashed lines represent the backward case.
}
\end{figure}

\begin{figure}[htb]
\includegraphics[width=6.5cm]{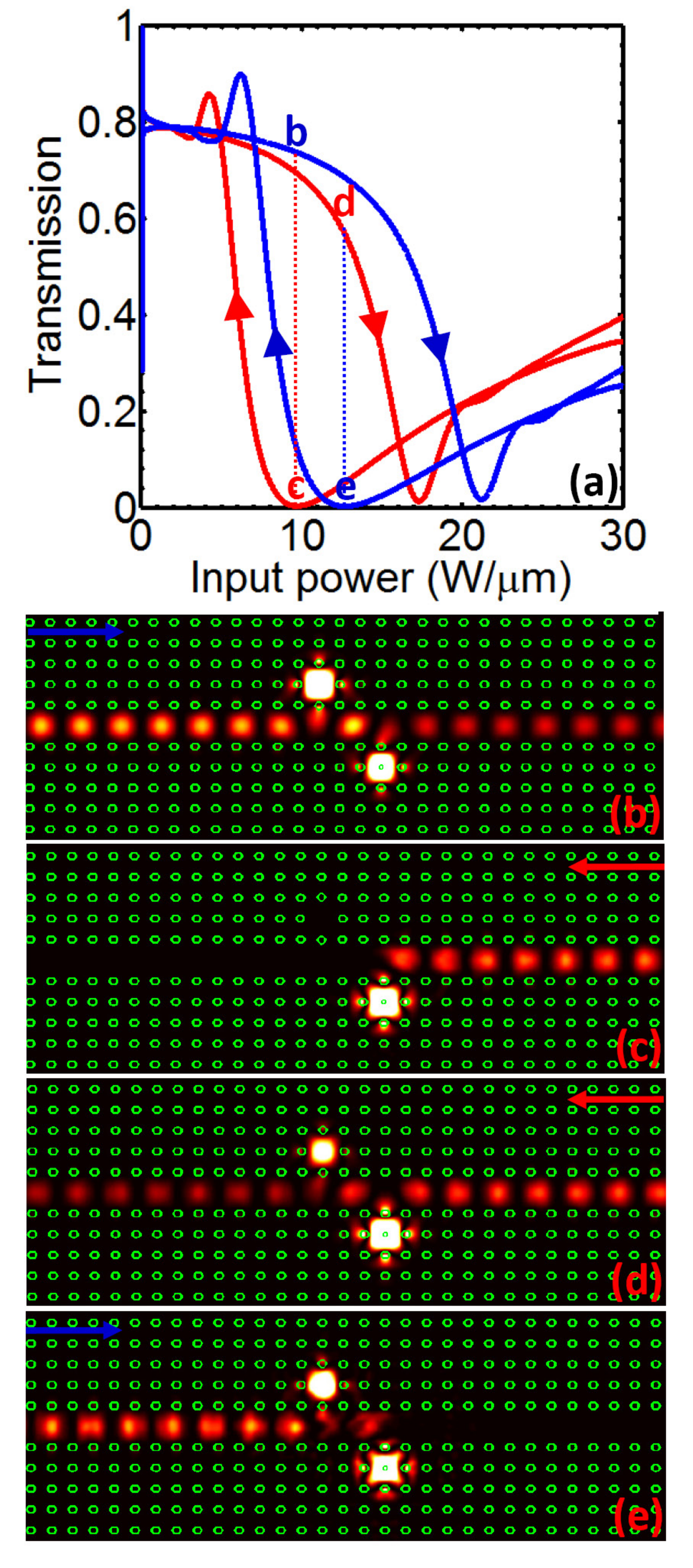}
\caption{\label{fig:fig6} 
(Color online) (a) The FDTD results of the nonlinear transmission at reversal incident directions. A Gaussian pulse with a carrier frequency $ f=0.378$ $ $ $2\pi c/a$ and $ 20 $ ps duration is used as an input probe signal. (b)-(e) are the corresponding instantaneous electric field $ \vert E\vert^{2} $ distribution marked b-e in (a), respectively. The structure details of the PhCs are outlined by the green lines. The side-coupled cavity on the right is with Kerr nonlinearity. (b)-(e) are normalized with each maximum and is saturated for better visualization. The FDTD grid is nonuniform. The grid size near the defect is $ a/100 $ to tell the small feature while the others are $  a/50 $. The computation domain is surrounded by perfect matching layers to absorb outgoing wave.
}
\end{figure}

\begin{figure*}[!t]
\includegraphics[width=2\columnwidth]{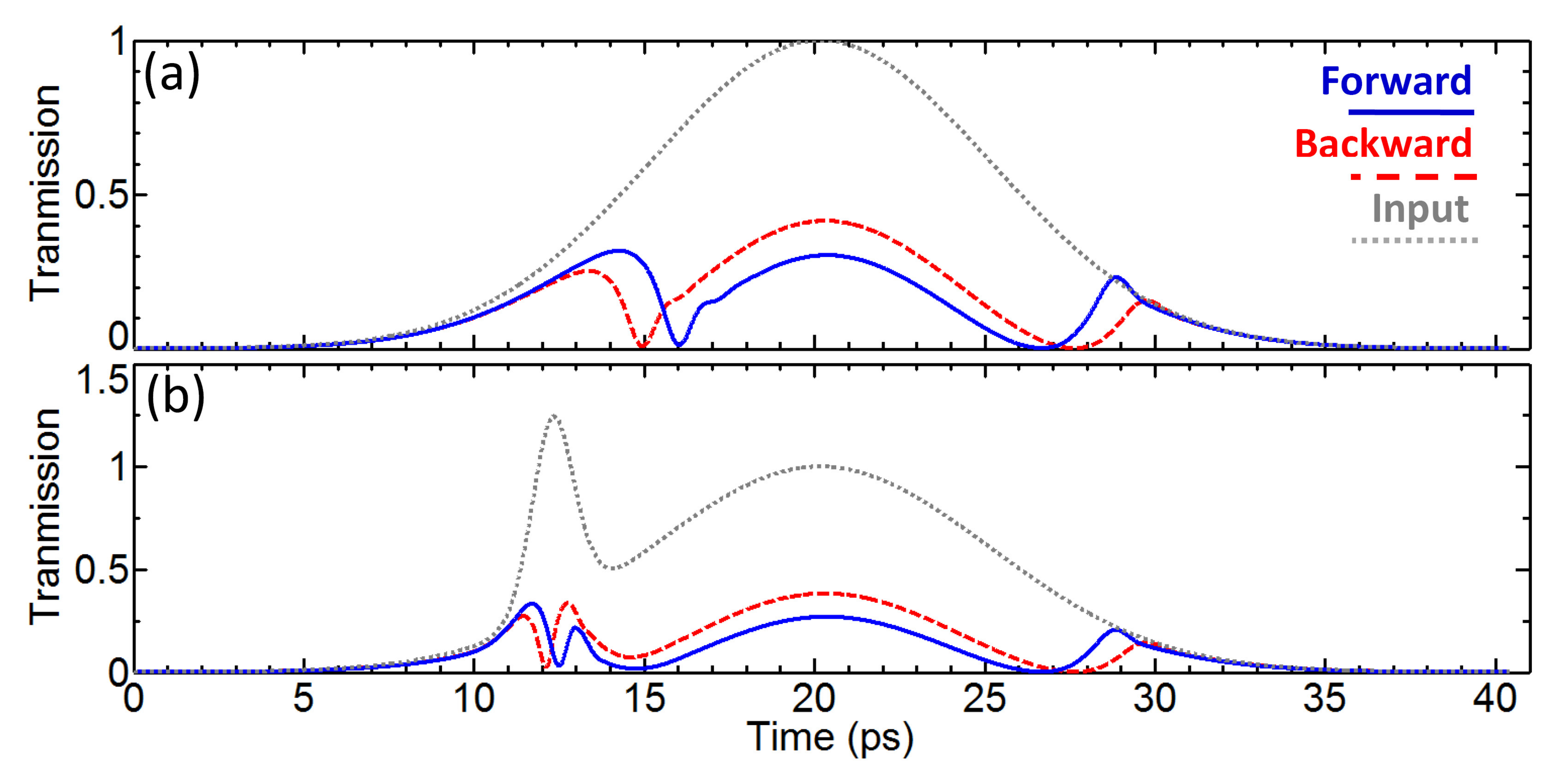}
\caption{\label{fig:fig7} 
(Color online) (a) The FDTD results of the pulse dynamics at reversal incident directions ($ \omega=0.378 2\pi c/a  $). The input signal, which is used as a probe pulse to access the bistability, is shown by the dotted line. (b)Reconfigurable nonreciprocity driven by a pump pulse whose duration is comparable to the nonlinear cavity's lifetime. The maximum amplitude of the pump pulse is one third of the probe one. The pump pulse is superimposed upon the input Gaussian pulse, as is shown by the dotted line in (b).
}
\end{figure*}

It is demonstrated that nonlinear Fano defect is suffered from modulational instability near the resonance under continuous-wave excitation~\cite{dynamic}, which resemble the scenario of wave scattering by a nonlinear center first addressed in Ref.~\cite{mi}. The modulation instability of our system is similar to the previous reported one~\cite{dynamic}. So, there is still a way to access the bistable hysteresis loop by a Gaussian pulses with suitable duration. Crank-Nicolson method~\cite{CK} with absorption boundary condition~\cite{dtbc} are used to solve Eqs. 1. Fig.~\ref{fig:fig5} (a) and (c) present the dynamics of the system under the excitation of a pulse where we only show part of pulse for better visualization. As can be seen from the marked forward directions of the diode in Fig.~\ref{fig:fig5} (a), the unidirectionality of the nonreciprocity is dynamically reconfigured at different time of the pulse's rising edge and falling edge, respectively. It is because the threshold power, which triggers the transition between two stable states at the cases of the increasing and decreasing input power, are both different referring to reversal incident wave, as are shown in Fig.~\ref{fig:fig5} (b). They valid that DRN could be realized by shaping the excitation condition to switch between the high-transmission states and the nonlinear Fano resonant low-transmission states. Fig.~\ref{fig:fig5} (b) also provides the analytical results (thick curves) which agrees with the dynamical simulations(thin curves). The oscillation between two bistable states indicates the transition between them and it is the property of the dynamical bistability~\cite{dynamic}. Fig.~\ref{fig:fig5} (c) and (d) present the dynamics of the power in the nonlinear Fano defect and the corresponding tunable power in nonlinear cavity, which demonstrates the incident direction dependence of the pumping rate in the nonlinear Fano defect. It is consistent with the idea of which we use directional dependent nonlinear Fano resonance to manipulate the wave transmission.

\section{Photonic crystal realization}
Now we provide a realistic example of reconfigurable nonreciprocity which is consisted of a linear photonic crystal (PhC) waveguide, with a side-coupled linear and nonlinear cavities pair. The PhC is formed by dielectric rods arranging in square lattice. The radius of the rods are $r=0.18a $, where $ a $ is the lattice constant and the refractive index $ n=3.4 $. Air is the background medium. The waveguide is created by removing one row of rods. And the nonlinear defect on the right is made by replacing one rod with a polymer rod with $r_{d}=0.1a $, $ n_{l}=1.6 $, $ n_{2}=1.14\times 10^{-12} cm^{2}/W $ and $ n=n_{l}+n_{2}I $, while the linear defect is introduced by removing one dielectric rod. The nearest rods of the linear defect are shifted $ 0.028a $ away from the center to keep the eigen-resonance frequencies of two defects the same and thus obtain flat stop band with sharp edge. The structure details are outlined by the green lines in Fig.~\ref{fig:fig6} (b)-(e). It was demonstrated that the nonlinear dynamics of this photonic structure can be described by a discrete model similar to the one presented in Sec. II~\cite{discrete}.

The finite different time domain (FDTD) simulation results [see Fig.~\ref{fig:fig6} (a) and Fig.~\ref{fig:fig7} (a)] obtained by a Gaussian pulse, show good agreement with the theoretical model. They prove that we can dynamically manipulate the nonreciprocity of the system by using the rising and falling edges of a pulse with suitable duration. The flipping of unidirectional transmission are indicated by b,c and d,e in Fig.~\ref{fig:fig6} (a), in which the forward direction of the Fano diode is reverse. We can use pump-assisting method to access low transmission case c and e marked in Fig.~\ref{fig:fig6} (a)~\cite{fan,lin}. The corresponding instantaneous electric field distributions $ \vert E\vert^{2} $ in the dynamic manipulation of the DRN are shown in Fig.~\ref{fig:fig6} (b)-(e). They manifest themselves as distinct interfering effects because different modes of the system are excited. For the Fano diode blocking the forward propagating wave, the interaction between two Fano defects is crucial [See Fig.~\ref{fig:fig6} (e)] while the nonlinear Fano defect dominates the resonant reflection when the forward direction is flipped [See Fig.~\ref{fig:fig6} (c)]. FDTD modelling detail can be found in the caption of Fig.~\ref{fig:fig6}. Furthermore, launching a suitable pump pulse (duration is comparable to the nonlinear cavity's lifetime) together with the probe one, it is possible to change the properties of the nonreciprocity. Because of distinct nonlinear feedbacks obtained at reversal excitations, their corresponding transmissions drop at different time referring to the same pump pulse, as is shown in Fig.~\ref{fig:fig7} (b). We can conclude from Fig.~\ref{fig:fig6} and Fig.~\ref{fig:fig7} that for the reversal incident waves, the nonreciprocity of the nonlinear Fano resonance system has distinct scenarios with respect to the input signal, which open up the possibility of dynamical wave manipulation. 

\section{Conclusions}
We suggested a concept of the dynamical reconfigurable nonreciprocity based on a nonlinear Fano resonance system. Tunable rectification, which manifests itself via tunable bistability, and dynamical reconfiguration of diode's forward direction is theoretically investigated. The interaction between the linear and nonlinear Fano resonances plays an important role in manipulating the nonreciprocity. Numerical experiments confirm our motivation via a realistic photonic example. Our results could pave the way for the advanced manipulation of wave rectification. Because of the similarity between the modified Fano-Anderson model and the discrete nonlinear Schr$ \ddot{o} $dinger equation which is shown to be relevant in many other physical contexts, we believe our results can be generalized to similar physical system such as the coupled optical waveguide arrays system~\cite{photo_wave_array} and many others.

\section*{ACKNOWLEDGEMENT}
The work of A. E. Miroshnichenko was supported by the Australian Research Council through Future Fellowship program (FT110100037). Y. Xu acknowledges the support from the National Natural Science Foundation (Grant No.11304047), Natural Science Foundation of Guangdong Province (Grant No.S201310014807) and the Xin Miao Science Foundation of Guangzhou University.

\end{document}